\documentclass[a4paper,journal]{IEEEtran}
\usepackage{amsmath,amsfonts}
\usepackage{algorithmic}
\usepackage{algorithm}
\usepackage{array}
\usepackage[caption=false,font=normalsize,labelfont=sf,textfont=sf]{subfig}
\usepackage{textcomp}
\usepackage{stfloats}
\usepackage{url}
\usepackage{verbatim}
\usepackage{graphicx}
\usepackage{cite}
\usepackage{multirow}
\usepackage{multicol}
\usepackage{booktabs}
\begin{document}

\title{Buffer-Parameterized Machine Learning Surrogate Models for Cross-Technology Signal Integrity Analysis and Optimization}

\author{Julian Withöft, Werner John, Emre Ecik, Ralf Brüning, Jürgen Götze
\thanks{This work has been submitted to the IEEE for possible publication. Copyright may be transferred without notice, after which this version may no longer be accessible.}%
\thanks{This work is funded as part of the research project KI4BoardNet in the funding programme MANNHEIM (BMBF) (Grant numbers 16ME0779/TUDO - 16ME0777/IDMT). The responsibility for this publication is held by the authors only.}%
\thanks{Julian Withöft, Emre Ecik and Jürgen Götze are with the Information Processing Lab, Faculty for Electrical Engineering and Information Technology, TU Dortmund, Germany (e-mail: \{julian.withoeft, emre.ecik, juergen.goetze\}@tu-dortmund.de)}%
\thanks{Werner John is with Pyramide2525, Paderborn, Germany and the Information Processing Lab, Faculty for Electrical Engineering and Information Technology, TU Dortmund, Germany (e-mail: werner.john@pyramide2525.de)}%
\thanks{Ralf Brüning is with EMC Technology Center Paderborn, Zuken GmbH, Paderborn, Germany (e-mail: ralf.bruening@de.zuken.com)}%
}

\maketitle

\begin{abstract}
Signal integrity (SI) analysis in printed circuit board (PCB) interconnects faces increasing complexity due to diverse integrated circuit (IC) buffer technologies, varying operating conditions, and manufacturing tolerances. Existing machine learning (ML) surrogate models for the prediction of SI metrics such as the inner eye contour, eye-height (EH), eye-width (EW), and transient waveform features often rely on fixed buffer parameters, limiting the adaptability across design and technology variations. This paper introduces a buffer-parameterized ML surrogate modeling methodology capable of handling cross-technology variations without retraining by treating IC buffer characteristics, e.g., clock frequency, supply voltage, rise/fall times, jitter, and internal resistors and capacitors, as model inputs in addition to the PCB parameters. To identify the optimal surrogate architecture for this high-dimensional space, tree-based methods (RFR/GBM), kernel methods (SVR/KRR), Gaussian process regression (GPR), and neural networks are compared. The buffer-parameterized approach is validated using a complex interconnect with a total of 44 design parameters. In addition to demonstrating strong generalization, the results show that neural networks outperform other ML models on large datasets, while the anisotropic GPR excels in low-data regimes. Finally, the practical value of the ML surrogate models is demonstrated through a cross-technology design space exploration and optimization scenario, which indicates large computational speedups for eye mask compliance checking compared to simulation.
\end{abstract}


\section{Introduction}
\label{sec_introduction}

\IEEEPARstart{M}{odern} electronic systems are characterized by increasing signal integrity (SI) requirements driven by higher clock rates, lowered voltage levels, and rise/fall times in the region of a few picoseconds. These trends significantly elevate the risk of signal reflections and distortions, making SI compliance a critical yet challenging aspect of printed circuit board (PCB) design. The complex interaction of parasitic physical couplings requires careful consideration of PCB trace configurations, routing topologies, and termination strategies. The resulting design rules and requirements have become incredibly tight, complex, and time-consuming to manage, often necessitating multiple design iterations to achieve adequate solutions.

Recent advances in artificial intelligence (AI) and machine learning (ML) have demonstrated promising potential to address these challenges by supporting PCB design in terms of SI analysis and optimization. The fundamental premise of AI/ML integration in this domain is to guide PCB designers towards appropriate design decisions through systematic capture and transfer of domain knowledge. This is achieved by training ML models on simulation data to enable prediction of various SI performance metrics such as transient waveform features such as max. overshoot and rise-time degradation as well as eye diagram features like the eye contour, eye-height (EH), and eye-width (EW).

By leveraging AI/ML-based systems that learn from extensive simulation data, it becomes possible to model the complex SI behavior and provide advanced design guidance. This approach promises to maintain design flexibility while effectively reducing design cycles and supporting SI-compliant design decisions.

The prediction of EH and EW from design parameters represents the most extensively studied area. Early work by \cite{Wu2018} employed an ML modeling approach to map design parameters of a PCB interconnect directly to these eye diagram metrics. This study compared artificial neural network (ANN), support vector regression (SVR), and linear regression models, finding that the ANN regression performed best due to its superior ability to handle the complex non-linear relationship. The SVR was just slightly behind the ANN in terms of performance, while the linear regression could not handle the nonlinearity of the task as expected. Further SVR applications for EH/EW prediction were implemented in \cite{Trinchero2018a,Ma2020}.

More advanced neural network implementations have incorporated transfer learning to enable model reuse across different environments \cite{Zhang2019} or semi-supervised learning \cite{Chen2020} and channel-characteristic modeling approaches \cite{Lho2022a} for improved prediction accuracy. In \cite{Nguyen2023}, Gaussian Process Regression (GPR) models with bounded EH and EW prediction using a beta likelihood and variational inference have been developed showing improved performance compared to an exact Gaussian implementation. With a focus on design optimization, a genetic algorithm (GA) was employed in \cite{Zhang2022} based on a trained ANN surrogate model for EH maximization. Similarly, \cite{Withoeft2024,Withoeft2025} also utilized a GA-ANN combination for design optimization to elaborate the feasible region of design parameters that fulfill EH/EW requirements under consideration of manufacturing tolerances, intra-pair skew, and common-mode restrictions.

\begin{figure*}[t!]
	\centering
	\includegraphics[width=17cm]{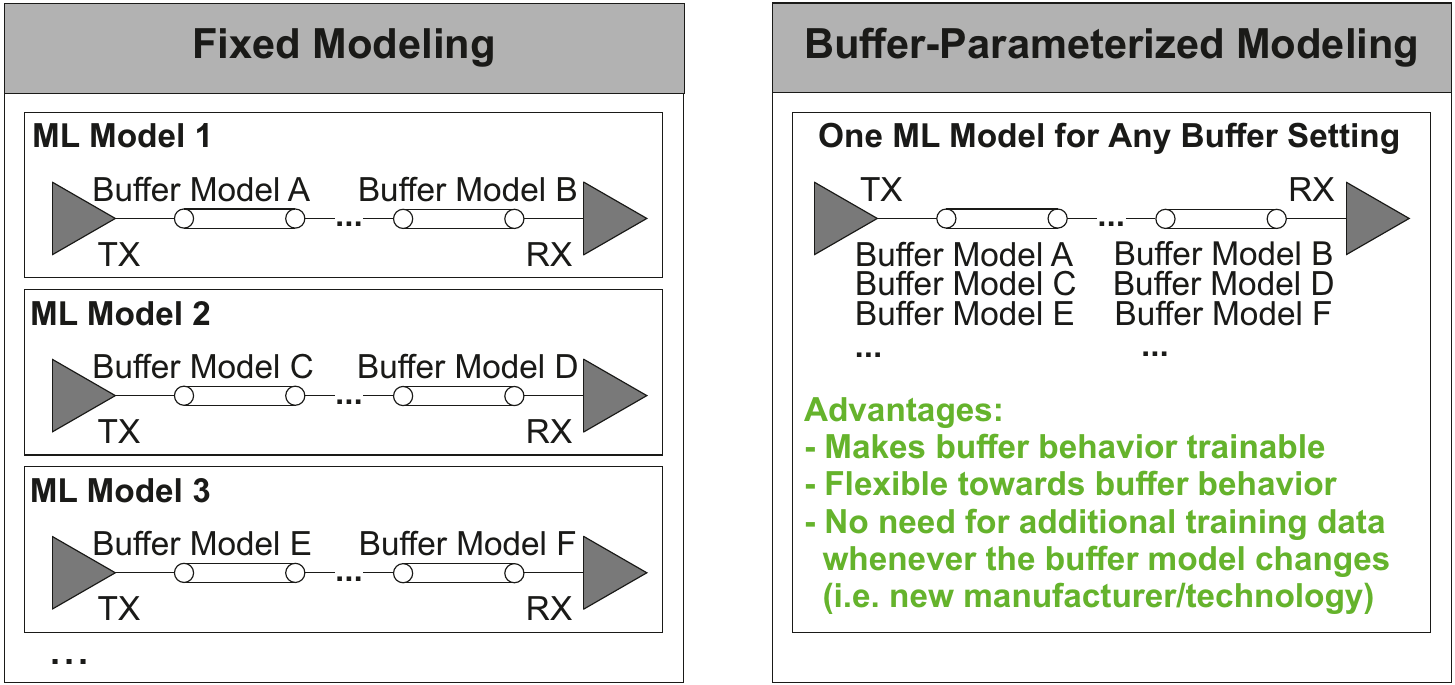}
	\caption{Fixed-buffer vs. buffer-parameterized ML modeling for SI analysis.}
	\label{fig_generalizable_approach}
\end{figure*}

Another eye-diagram-based ML modeling approach is the prediction of the entire inner eye contour from design parameters instead of just the EH and EW metrics. Initially, ANN models were utilized in \cite{Goay2019} for the prediction of the eye contour based on a simple point-to-point interconnect by variation of the length and impedance through changes in geometry. The approach specifically focused on adaptive sampling to improve the eye contour prediction performance. The approach in \cite{Telescu2023b} simplified the problem by predicting the eye area that is enclosed by the contour as a single parameter with an SVR. Then, in \cite{Telescu2023a}, the prediction of the entire eye contour was realized using SVR and kernel ridge regression (KRR) models indicating that on average the KRR provides more accurate predictions than the SVR.

Beyond eye diagram analysis, transient waveform metrics such as energy, entropy, rise time, and overshoot have been modeled using ML surrogate approaches in \cite{Withoeft2023a}. To provide PCB designers with explainable AI models, decision trees (DTs) were introduced for SI anomaly detection in \cite{Ecik2023a} and combined with SVR to systematically extract entire SI-compliant parameter spaces \cite{Ecik2024b}. In addition to these anomaly detection methodologies, transfer learning principles have been applied in \cite{Withoeft2023b} to improve data efficiency for a technology change from DDR3 to DDR4, while reinforcement learning (RL) was utilized in \cite{Withoeft2023c} for design optimization.

Other modeling efforts, such as the work in \cite{Penugonda2020}, have successfully used ML to map transmission line geometries to frequency-dependent RLGC parameters. This approach replaces time-consuming electromagnetic field solvers for transmission line characterization, but still requires a circuit simulation step to obtain the final SI metrics. Direct modeling approaches map design parameters straight to SI metrics and avoid this intermediate circuit simulation step, which results in significant computational speedups required for extensive pre-layout design space exploration and optimization.

\begin{table}[b!]
	\caption{Overview of utilized ML models and their abbreviations}
	\centering
	\label{tab_model_abbreviations}
	\begin{tabular}{ll}
		\toprule
		\textbf{Machine Learning Surrogate Model} & \textbf{Abbreviation} \\
		\midrule
		Fully-Connected Artificial Neural Network & ANN \\
		Ensemble Fully-Connected Artificial Neural Network & Ens. ANN \\
		Encoder-Decoder Neural Network & Enc-Dec \\
		Ensemble Encoder-Decoder Neural Network & Ens. Enc-Dec \\
		Random Forest Regression & RFR \\
		Gradient Boosting Machine & GBM \\
		Support Vector Regression (Isotropic) & SVR-ISO \\
		Kernel Ridge Regression (Isotropic) & KRR-ISO \\
		Kernel Ridge Regression (Anisotropic) & KRR-ANI \\
		Exact Gaussian Process Regression (Isotropic) & GPRE-ISO \\
		Exact Gaussian Process Regression (Anisotropic) & GPRE-ANI \\
		Approx. Gaussian Process Regression (Isotropic) & GPRA-ISO \\
		Approx. Gaussian Process Regression (Anisotropic) & GPRA-ANI \\
		\bottomrule
	\end{tabular}
\end{table}

\begin{figure}[b!]
	\centering
	\includegraphics[width=8.5cm]{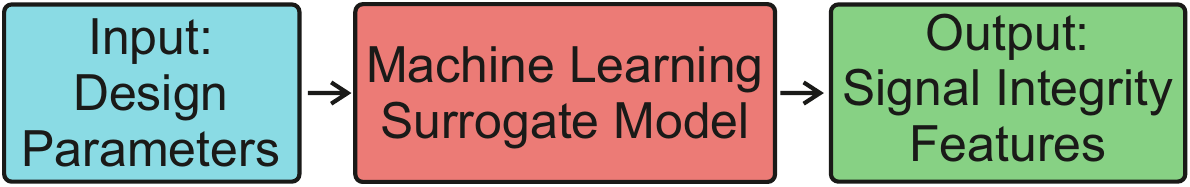}
	\caption{Conceptual workflow of ML surrogate modeling for predicting SI features at the output from design parameters at the input.}
	\label{fig_ml_modeling}
\end{figure}

While the previous studies have demonstrated that ML is a promising approach for SI analysis and design, they all share a key methodological limitation: the portability and generalizability with respect to varying integrated circuit (IC) buffer behavior. State-of-the-art approaches typically train the ML models by varying the PCB parameters and capturing the SI behavior at the receiver with a surrogate model. This works well as long as the buffer characteristics (e.g., driver strength, capacitance, rise/fall times) remain fixed. However, once these characteristics change or when a different technology with altered voltage levels or clock frequencies is introduced, the trained ML model becomes obsolete, as it has never been exposed to these kinds of operating conditions during training. This results in repeated cycles of data generation and retraining each time the IC buffer behavior changes, which is inefficient and cumbersome. This is also illustrated in Figure \ref{fig_generalizable_approach} with the fixed-buffer modeling approach, where there are separate models for each buffer configuration. While transfer learning is one way to reduce this recurring simulation effort, it still does not entirely alleviate this burden.

To overcome this inefficiency, this paper proposes a fundamental methodological shift in the form of a buffer-parameterized surrogate framework, which describes IC buffer behavior through the ML surrogate model's input parameters and thereby makes it trainable. In high-speed design workflows, PCB designers often reuse a limited library of interconnect topologies, while the primary variables of different product generations are the transmission line lengths, operating speeds, and IC buffer characteristics \cite{Withoeft2024}. With the buffer-parameterized framework, the simulation data only needs to be generated once and the ML model trained on this data is then applicable to any occurring buffer specifications within the range of the trainable IC buffer parameters. This removes the simulation completely from the optimization loop and allows for accelerated cross-technology SI optimization and design space exploration. Hence, the main contributions of this work are to:

\begin{itemize}
	\item Introduce a buffer-parameterized modeling methodology, which treats IC buffer characteristics as trainable input parameters to break the retraining cycle of fixed buffer models.
	\item Conduct a benchmarking study of various regression methods as shown in Table \ref{tab_model_abbreviations} and Figure \ref{fig_ml_modeling} as well as varying training dataset sizes to identify the optimal surrogate model type for this high-dimensional space.
	\item Assess the scalability and complexity of the proposed approach with respect to the number of design parameters using a highly complex scenario with a total of 44 design parameters to demonstrate the practical efficiency for multi-generational DDR design space exploration and optimization.
\end{itemize}

\section{Interconnect Topologies and Data Generation}

\begin{table}[b!]
	\caption{Parameter ranges for the microstrip line of the simple fixed-buffer point-to-point topology shown in Fig.~\ref{fig_simple_topology}}
	\centering
	\label{tab_simple_parameters}
	\begin{tabular}{lcc}
		\toprule
		\textbf{Parameter} & \textbf{Min} & \textbf{Max} \\
		\midrule
		Substrate Height ($h$) [mm] & 0.1 & 0.5 \\
		Trace Width ($w$) [mm] & 0.1 & 0.7 \\
		Trace Thickness ($t$) [mm] & 0.01 & 0.1 \\
		Relative Permittivity ($\epsilon_r$) & 3 & 5 \\
		Transmission Line Length ($l$) [mm] & 10 & 100 \\
		\bottomrule
	\end{tabular}
\end{table}

To demonstrate and compare the buffer-parameterized approach, three different PCB structures were considered.

\begin{figure}[b!]
	\centering
	\includegraphics[width=8.5cm]{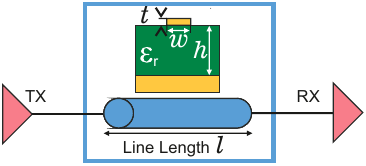}
	\caption{Simple fixed-buffer point-to-point topology.}
	\label{fig_simple_topology}
\end{figure}

\begin{table}[b!]
	\caption{Parameter ranges for the complex point-to-point topology shown in Fig.~\ref{fig_complex_topology}}
	\centering
	\label{tab_complex_parameters}
	\begin{tabular}{lcc}
		\toprule
		\textbf{Parameter [Unit]} & \textbf{Min} & \textbf{Max} \\
		\midrule
		Clock Frequency $f_{\text{clock}}$ [MHz] & 800 & 2400 \\
		Rise/Fall Times [\% of $T_{\text{clock}}$] & 0 & 40 \\
		Supply Voltage Level [V] & 0.8 & 1.8 \\
		Jitter [\% of $T_{\text{clock}}$] & 0 & 5 \\
		TX Driver Resistance $R_{\text{in,tx}}$ [$\Omega$] & 10 & 50 \\
		TX Driver Capacitance $C_{\text{in,tx}}$ [pF] & 0.1 & 2.0 \\
		RX Capacitance $C_{\text{in,rx}}$ [pF] & 0.1 & 2.0 \\
		\midrule
		TL1 Length $l_{\text{TL1}}$ [mm] & 1 & 100 \\
		TL2 Length $l_{\text{TL2}}$ [mm] & 1 & 10 \\
		TL3 Length $l_{\text{TL3}}$ [mm] & 1 & 100 \\
		TL4 Length $l_{\text{TL4}}$ [mm] & 0.1 & 10 \\
		TL5 Length $l_{\text{TL5}}$ [mm] & 0.1 & 10 \\
		TL6 Length $l_{\text{TL6}}$ [mm] & 0.1 & 10 \\
		\midrule
		TL1 Characteristic Impedance $Z_{\text{0,TL1}}$ [$\Omega$] & 30 & 90 \\
		TL2 Characteristic Impedance $Z_{\text{0,TL2}}$ [$\Omega$] & 30 & 90 \\
		TL3 Characteristic Impedance $Z_{\text{0,TL3}}$ [$\Omega$] & 30 & 90 \\
		TL4 Characteristic Impedance $Z_{\text{0,TL4}}$ [$\Omega$] & 30 & 90 \\
		TL5 Characteristic Impedance $Z_{\text{0,TL5}}$ [$\Omega$] & 30 & 90 \\
		TL6 Characteristic Impedance $Z_{\text{0,TL6}}$ [$\Omega$] & 30 & 90 \\
		\midrule
		TL1 Dielectric $\epsilon_{\text{eff,TL1}}$ & 3 & 5 \\
		TL2 Dielectric $\epsilon_{\text{eff,TL2}}$ & 3 & 5 \\
		TL3 Dielectric $\epsilon_{\text{eff,TL3}}$ & 3 & 5 \\
		TL4 Dielectric $\epsilon_{\text{eff,TL4}}$ & 3 & 5 \\
		TL5 Dielectric $\epsilon_{\text{eff,TL5}}$ & 3 & 5 \\
		TL6 Dielectric $\epsilon_{\text{eff,TL6}}$ & 3 & 5 \\
		\midrule
		Termination Resistor $R_{\text{T}}$ [$\Omega$] & 1 & 100 \\
		\midrule
		TX Package Resistance $R_{\text{pkg,tx}}$ [m$\Omega$] & 10 & 500 \\
		TX Package Inductance $L_{\text{pkg,tx}}$ [nH] & 0.5 & 5.0 \\
		TX Package Capacitance $C_{\text{pkg,tx}}$ [pF] & 0.1 & 2.0 \\
		\midrule
		RX Package Resistance $R_{\text{pkg,rx}}$ [m$\Omega$] & 10 & 500 \\
		RX Package Inductance $L_{\text{pkg,rx}}$ [nH] & 0.5 & 2.0 \\
		RX Package Capacitance $C_{\text{pkg,rx}}$ [pF] & 0.1 & 2.0 \\
		\midrule
		TX Via Resistance $R_{\text{via,tx}}$ [m$\Omega$] & 0.1 & 5.0 \\
		TX Via Inductance $L_{\text{via,tx}}$ [nH] & 0.1 & 2.0 \\
		TX Via Capacitance 1 $C_\text{{via1,tx}}$ [pF] & 0.05 & 0.8 \\
		TX Via Capacitance 2 $C_\text{{via2,tx}}$ [pF] & 0.05 & 0.8 \\
		\midrule
		RX Via Resistance $R_\text{{via,rx}}$ [m$\Omega$] & 0.1 & 5.0 \\
		RX Via Inductance $L_\text{{via,rx}}$ [nH] & 0.1 & 2.0 \\
		RX Via Capacitance 1 $C_\text{{via1,rx}}$ [pF] & 0.05 & 0.8 \\
		RX Via Capacitance 2 $C_\text{{via2,rx}}$ [pF] & 0.05 & 0.8 \\
		\midrule
		Termination Via Resistance $R_\text{{via,T}}$ [m$\Omega$] & 0.1 & 5.0 \\
		Termination Via Inductance $L_\text{{via,T}}$ [nH] & 0.1 & 2.0 \\
		Termination Via Capacitance 1 $C_\text{{via1,T}}$ [pF] & 0.05 & 0.8 \\
		Termination Via Capacitance 2 $C_\text{{via2,T}}$ [pF] & 0.05 & 0.8 \\
		\bottomrule
	\end{tabular}
\end{table}

Initially, for demonstration of the fixed-buffer modeling approach, the structure shown in Fig.~\ref{fig_simple_topology} was utilized, which is a simple topology consisting of only a single microstrip transmission line connecting driver (TX) and receiver (RX). This is an example of a fixed-buffer modeling scenario, consistent with the literature discussed in Section \ref{sec_introduction}, as the driver and receiver parameters were fixed and only the parameters of the microstrip were varied according to Table \ref{tab_simple_parameters}. Eye diagram simulations were carried out based on design of experiments (DoE) using individual Latin hypercube sampling (LHS) for training (4,000 samples), validation (1,000 samples), and test (1,000 samples) datasets to obtain EH, EW, and the eye contour \cite{LTspice}.

In the second scenario, for investigation of the buffer-parameterized approach, the point-to-point structure and parameter variation of \cite{Withoeft2025} were utilized. The IC buffer behavior of the RX and TX was defined as variable in the sense that besides the PCB parameters e.g., transmission line lengths, characteristic impedances, dielectric constant, and termination resistance, the clock frequency dictating the datarate, supply voltage level, jitter, rise/fall times as well as internal resistance and capacitance values were also varied. A total of 40,000 training, 10,000 test, and 10,000 validation samples were obtained for eye diagram and transient waveform simulations based on LHS-DoE \cite{LTspice}. Again, EH/EW and eye contour were extracted from the eye diagram simulations, and additionally, the waveform features energy, entropy, maximum voltage, minimum voltage, rise time, overshoot, and propagation delay were extracted from the transient waveform simulations.

\begin{figure*}[t!]
	\centering
	\includegraphics[width=17cm]{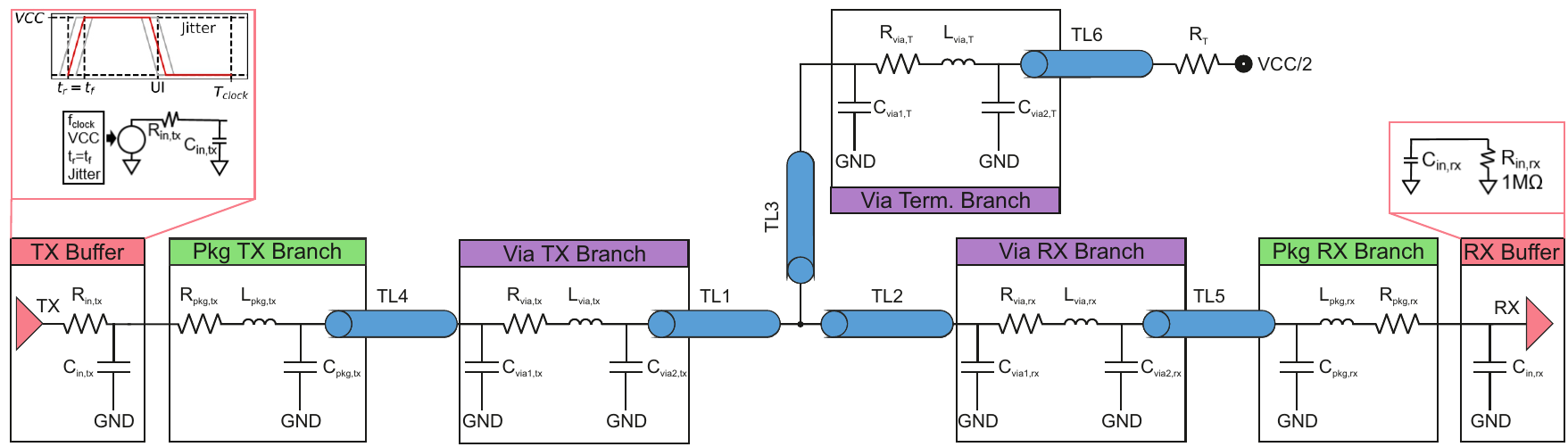}
	\caption{Complex point-to-point topology including package parasitics and via parasitics.}
	\label{fig_complex_topology}
\end{figure*}

Finally, for a more complex scenario, the interconnect shown in Fig.~\ref{fig_complex_topology} was utilized. This is based on a point-to-point topology, but with added IC package parasitics and vias in the interconnect path, which were parameterized based on their RLC parameters as shown in Table \ref{tab_complex_parameters}. The specified design parameter ranges represent typical values derived from various IBIS models and IC datasheets. Moreover, there were more transmission line segments in the path, and the phase velocity of each segment was individually characterized by $\epsilon_{\text{eff}}$. While the proposed framework can incorporate loss tangent as an additional trainable feature, it was kept fixed across all investigated scenarios in this study, as dielectric attenuation represents a secondary effect compared to the mismatch reflections dominating these specific topologies. The clock frequency was also adjusted to be significantly higher in the range of 800-2400 MHz to evaluate the method at even higher speeds. This significantly extends the number of design parameters to a total of 44 and can be seen as an example of a realistic and complex topology, which is parameterizable and adaptable with respect to all relevant design parameters. Due to the high parameter count, 100,000 training samples were generated, while the test and validation datasets remained the same size as before (10,000) \cite{LTspice}.

In general, the EH and EW measurements presented in this work are based on a modified contour-based methodology, which differs from the statistical approach used in \cite{Withoeft2024,Withoeft2025}. In the previous studies \cite{Withoeft2024,Withoeft2025}, EH and EW were computed statistically by subtracting three standard deviations (3$\sigma$) from the mean of the high and low voltage levels and the crossing times, respectively. In contrast, the modified method directly measures the absolute worst-case minimum vertical and horizontal opening of the eye diagram's contour. This inner boundary includes asymmetric, deterministic, and non-Gaussian waveform distortions that a purely statistical extrapolation might misrepresent. This process was realized using a custom algorithm that isolates the inner eye contour as a 100-dimensional vector with 50 samples for the upper and 50 for the lower contour based on which subsequently the specific EH and EW metrics are computed.

For pre-processing of the data, the input design parameters were normalized. EH and EW were also normalized according to the supply voltage level and unit interval (UI), respectively. The 100-dimensional eye contour vector was not processed, and the waveform features were standardized with mean and standard deviation according to \cite{Withoeft2023b}. The validation size was set to 25\% of the training data throughout all experiments to ensure comparability across data sizes and to align with the typical 80\%-20\% train-validation split. The number of testing samples remained unaffected to provide a common ground for comparison.

\section{Machine Learning Methods}

\subsection{Tree-Based Methods}

Tree-based methods such as decision tree regression (DTR), random forest regression (RFR), and gradient boosting machines (GBMs) represent a class of ML models that generate predictions through a series of hierarchical decisions, which forms a tree-like structure. They are baselines for tabular, non-linear parametric regression. However, their discrete splitting nature can limit performance for smooth and continuous waveform predictions.

DTR recursively partitions the feature space into rectangular regions by splitting on feature values. Each region corresponds to a leaf, where the model predicts a constant value equal to the mean of the target values of the training samples in that leaf.

RFR is essentially an ensemble of DTRs, each built on a random subset of the data and considering only a random subset of features at each split. This intentional randomness reduces the correlation between individual trees, which leads to a more robust and accurate overall prediction that is obtained by averaging the predictions from all the individual trees in the forest \cite{Breiman2001}.

GBMs, on the other hand, train decision trees sequentially and each new tree attempts to minimize the residuals from the predictions of all previously trained trees. This sequential approach focuses on instances with high prediction errors in previous stages to increase overall GBM accuracy by learning from its mistakes \cite{Friedman2001}.

\subsection{Gaussian Process Regression}

Gaussian processes (GPs) provide a robust Bayesian non-parametric framework for regression. A function $f(\mathbf{x})$ is drawn from a GP defined by its mean $m(\mathbf{x})$ and covariance (kernel) function $k(\mathbf{x}, \mathbf{x}')$:
\begin{equation}
	f(\mathbf{x}) \sim \mathcal{GP}(m(\mathbf{x}), k(\mathbf{x}, \mathbf{x}')).
\end{equation}

Observations are modeled as $\mathbf{Y} = \mathbf{f} + \boldsymbol{\epsilon}$, where $\boldsymbol{\epsilon} \sim \mathcal{N}(\mathbf{0}, \sigma_n^2 \mathbf{I})$ represents Gaussian noise. The kernel is essential to a GP as it models the similarity between function values at any two input points $\mathbf{x}$ and $\mathbf{x}'$. In this work, radial basis function (RBF) kernels are exclusively employed, with two primary forms considered:
\begin{itemize}
	\item \textbf{Isotropic RBF Kernel:} This form assumes a single lengthscale $\ell$ for all input dimensions, which implies that all features contribute equally to the similarity measure:
	\begin{equation}
	k(\mathbf{x}, \mathbf{x}') = \sigma_{\text{signal}}^2 \exp\left(-\frac{\|\mathbf{x} - \mathbf{x}'\|^2}{2\ell^2}\right),
	\end{equation}
	where $\sigma_{\text{signal}}^2$ is the signal variance.
	\item \textbf{Anisotropic RBF Kernel:} This more flexible form allows for separate lengthscales $\ell_d$ for each input dimension $d$ and therefore enables automatic relevance determination (ARD), where the model learns the individual relevance of each input feature:
	\begin{equation}
	k(\mathbf{x}, \mathbf{x}') = \sigma_{\text{signal}}^2 \exp\left(-\frac{1}{2} \sum_{d=1}^{D} \frac{(x_d - x'_d)^2}{\ell_d^2}\right).
	\end{equation}
\end{itemize}

For exact GPR \cite{Rasmussen2006}, the predictive mean $\mu(\mathbf{x}^*)$ and variance $\sigma^2(\mathbf{x}^*)$ for a test point $\mathbf{x}^*$ are given by $\mu(\mathbf{x}^*) = \mathbf{k}(\mathbf{x}^*, \mathbf{X}) (\mathbf{K}(\mathbf{X}, \mathbf{X}) + \sigma_n^2 \mathbf{I})^{-1} \mathbf{Y}$ and $\sigma^2(\mathbf{x}^*) = k(\mathbf{x}^*, \mathbf{x}^*) - \mathbf{k}(\mathbf{x}^*, \mathbf{X}) (\mathbf{K}(\mathbf{X}, \mathbf{X}) + \sigma_n^2 \mathbf{I})^{-1} \mathbf{k}(\mathbf{X}, \mathbf{x}^*)$, where $\mathbf{K}(\mathbf{X}, \mathbf{X})$ is the training kernel matrix, $\mathbf{k}(\mathbf{x}^*, \mathbf{X})$ is the covariance vector between $\mathbf{x}^*$ and training inputs, $\mathbf{k}(\mathbf{X}, \mathbf{x}^*)$ is its transpose, and $k(\mathbf{x}^*, \mathbf{x}^*)$ is the prior variance at the test point. This approach, however, quickly becomes intensive computationally and memory-wise due to the cubic computational scaling with the number of training samples ($O(N^3)$) required for matrix inversion, as well as the quadratic memory scaling ($O(N^2)$) to store the covariance matrix. This makes exact GP impractical for large datasets. Therefore, in this work, exact GP was only employed for lower dimensional tasks such as the discrete EH/EW (max. 5000 training samples) and waveform feature (max. 1000 training samples) modeling, but not for the eye contour prediction.

To address scalability to larger datasets, approximate variational GPR \cite{Titsias2009} is predominantly used. This approach introduces $M$ inducing points $\mathbf{Z}$ (where $M \ll N$) with corresponding function values $\mathbf{u}$ and transforms the optimization into maximization of the evidence lower bound (ELBO):
\begin{equation}
\mathcal{L}_{\text{ELBO}} = \mathbb{E}_{q(\mathbf{f})}[\log p(\mathbf{Y} | \mathbf{f})] - \text{KL}(q(\mathbf{u}) || p(\mathbf{u})),
\end{equation}
where $q(\mathbf{u}) = \mathcal{N}(\boldsymbol{\mu}_u, \mathbf{\Sigma}_u)$ is the variational posterior over the inducing points $\mathbf{u}$, and $q(\mathbf{f}) = \int p(\mathbf{f}|\mathbf{u}) q(\mathbf{u}) d\mathbf{u}$ is the marginal posterior over the function values $\mathbf{f}$. This reduces the computational complexity to $O(NM^2)$ and the memory requirement to $O(NM)$, making the approach tractable. In this work, the number of inducing points $M$ was set to $M = \min(1000, 0.5 \times N)$.

For multi-output GPs, the linear coregionalization model (LCM) \cite{Bonilla2007} describes each of the $Q$ output functions $f_q(\mathbf{x})$ as a linear combination of $L$ independent latent GPs:
\begin{equation}
f_q(\mathbf{x}) = \sum_{l=1}^{L} w_{ql} g_l(\mathbf{x}), \quad g_l(\mathbf{x}) \sim \mathcal{GP}(0, k_l(\mathbf{x}, \mathbf{x}')),
\end{equation}
where $w_{ql}$ denotes the mixing weight for latent $l$ and output $q$. In this work, $L=2$ for EH/EW, $L=7$ for waveform features, and $L=3$ for eye contour prediction was chosen to keep computations feasible with respect to all training sample sizes.

\subsection{Kernel Methods}

Kernel methods represent a class of ML algorithms that implicitly map input data into a higher-dimensional feature space, where linear models can then be applied. This transformation, often referred to as the kernel trick, allows the models to capture complex nonlinear relationships in the original input space without explicitly computing the coordinates in the higher-dimensional space \cite{Schoelkopf2002}. The core of these methods lies in the choice of the kernel function $k(\mathbf{x}, \mathbf{x}')$, here implemented using the same isotropic and anisotropic RBF kernels applied in the GPR models.

SVR adapts the principles of support vector machines (SVMs) to regression tasks. The goal of SVR is to find a function that deviates from the training data by no more than a specified margin $\epsilon$, while simultaneously being as flat as possible. It achieves this by defining an $\epsilon$-insensitive tube around the regression line, which penalizes only those errors that fall outside this margin. For multi-output regression, the MSVR implementation from \cite{Bao2014} was utilized, which is designed to handle multi-target prediction through a vector-valued formulation. For a new input $\mathbf{x}^*$, the MSVR model obtains a $Q \times 1$ prediction vector $\hat{\mathbf{y}}(\mathbf{x}^*)$ computed as
\begin{equation}
	\hat{\mathbf{y}}(\mathbf{x}^*) = \mathbf{A}^\top \mathbf{k}(\mathbf{x}^*, \mathbf{X})^\top + \mathbf{b},
\end{equation}
where $\mathbf{A}$ is the $N \times Q$ matrix of learned coefficients and $\mathbf{b}$ is the $Q \times 1$ column bias vector. Extensive hyperparameter tuning (1,000 iterations) failed to achieve satisfactory performance for the anisotropic SVR model, which indicated that the length scales could not be properly optimized. As a result, only the isotropic SVR predictions are presented.

KRR extends normal ridge regression using the kernel trick to implicitly map data into a higher-dimensional feature space. It aims to find a function that minimizes the sum of squared errors while simultaneously penalizing the complexity of the model using L2 regularization on the coefficients to prevent overfitting. A vector-valued KRR (VVKRR) implementation based on \cite{Soleimani2023} was employed, which provides multi-output support and modeling of dependencies between output dimensions. The prediction for a new input $\mathbf{x}^*$ is computed as
\begin{equation}
	\hat{\mathbf{y}}(\mathbf{x}^*) = \mathbf{B} \mathbf{C}^\top \mathbf{k}(\mathbf{x}^*, \mathbf{X})^\top,
\end{equation}
where $\mathbf{C}$ is the $N \times Q$ matrix of learned coefficients, and $\mathbf{B}$ is the $Q \times Q$ output kernel matrix, modeling the correlations between outputs.

\subsection{Neural Networks}

Artificial Neural Networks (ANNs) are among the most widely used ML architectures and are often referred to as fully-connected/dense networks or multi-layer perceptrons (MLPs). They are characterized by layers in which each neuron is connected to all neurons in the subsequent layer. This architecture allows ANNs to learn highly complex mappings from input to output, making them universal function approximators capable of modeling a wide range of functions \cite{Goodfellow2016}.

\begin{figure}[b!]
	\centering
	\includegraphics[width=8.5cm]{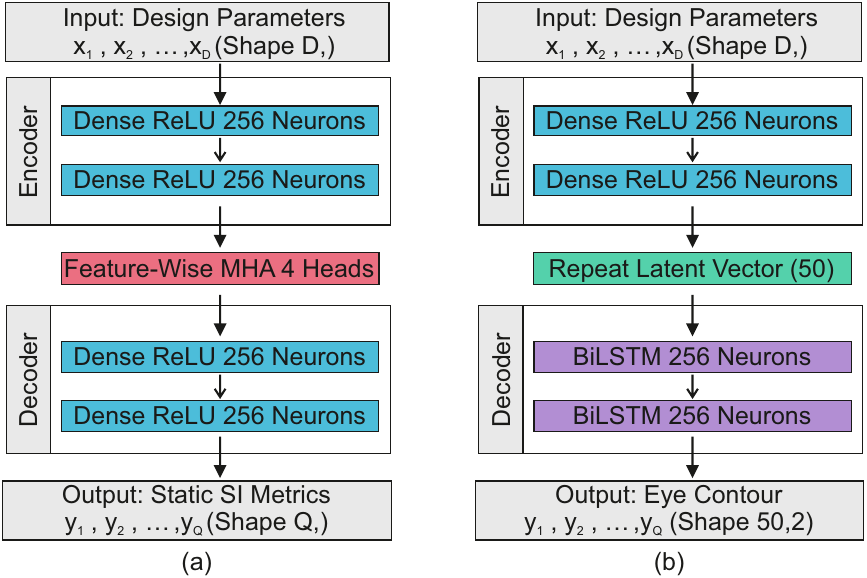}
	\caption{Encoder-decoder architectures for prediction of discrete SI metrics such as EH, EW, overshoot, and rise time (a) and the eye contour (b).}
	\label{fig_encoder_decoder}
\end{figure}

Besides standard fully-connected ANN-based models, more complex encoder-decoder neural network architectures \cite{Sutskever2014} are also considered in this work. An encoder-decoder neural network splits the prediction task into two parts: an encoder processes the input data i.e. the design parameters into a context vector and the decoder then generates the output from this context.

For discrete SI metrics such as EH, EW, overshoot, and rise time, the multi-head attention (MHA)-based dense encoder-decoder architecture shown in Fig.~\ref{fig_encoder_decoder} (a) was used. It is important to note that the MHA here operates as a feature-wise attention mechanism on the static input vector. It allows the model to weight the relevance of the high-dimensional design parameters when predicting scalar outputs. For the eye contour prediction, a dense encoder paired with a recurrent bidirectional long short-term memory (Bi-LSTM) decoder shown in Fig.~\ref{fig_encoder_decoder} (b) was employed. Eye contour prediction requires unfolding a static latent vector into a short 50-sample sequence. The Bi-LSTM decoder evaluates the latent representation of the design parameters at every decoding timestep. Furthermore, the eye contour represents a correlated and smooth continuous physical waveform. Therefore, attention mechanisms were intentionally omitted from the contour decoder. The experiments showed that for a sequence of this nature, the Bi-LSTM is sufficient and adding attention only led to unnecessary parameterization, increased computational overhead, and higher overfitting risks without achieving meaningful accuracy improvements.

To further improve predictive accuracy and robustness, ensemble learning was applied to the neural network models. This involved the training of the same neural network architecture with an ensemble size of 25 times independently. The predictions from these 25 individual networks were then averaged to obtain the final ensemble output. This strategy utilizes the diversity from random initializations and training variability, which leads to improved accuracy and stability compared to single models as demonstrated in \cite{Hansen1990}.

\subsection{Hyperparameter Optimization}

The predictive performance of ML models is deeply connected to model-specific hyperparameters, which dictate convergence and generalization. Since these parameters cannot be learned directly from the data, an efficient and automated search strategy is required. Bayesian optimization (BO) was employed, following methodologies established in prior studies \cite{Withoeft2023a,Withoeft2025}. This approach is implemented via the Optuna framework \cite{Optuna2019} and constructs a probabilistic surrogate model to prioritize high-potential regions of the search space. This results in better sample efficiency compared to grid or random searches. A computational budget of 100 iterations was allocated to ensure identification of optimal hyperparameters.

\begin{figure}[t!]
	\centering
	\includegraphics[width=8.5cm]{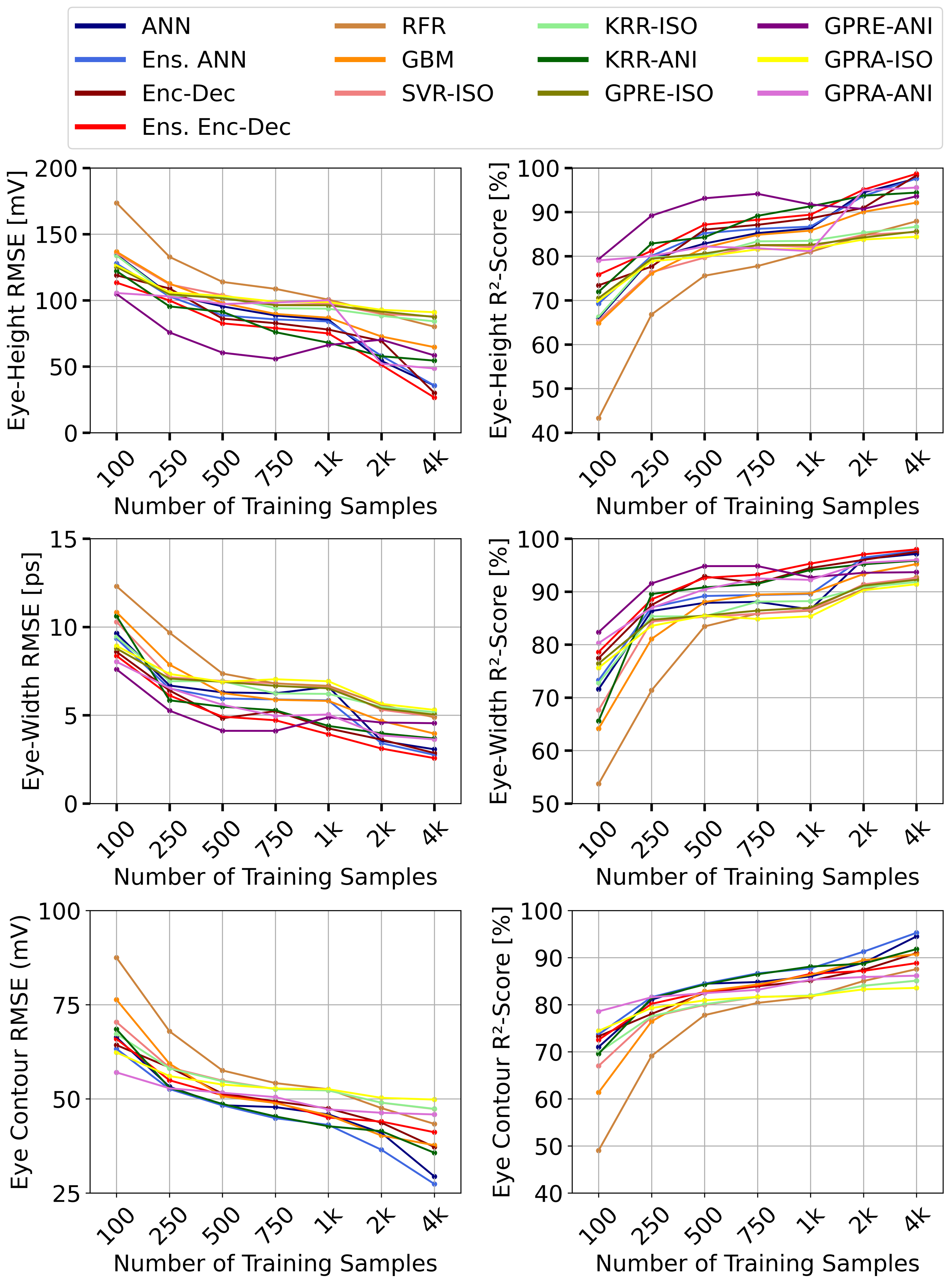}
	\caption{Testing RMSE and $R^2$-score for prediction of EH, EW, and eye contour for a varying number of training samples based on the simple fixed-buffer scenario of Fig.~\ref{fig_simple_topology}.}
	\label{fig_ml_results_simple_spice}
\end{figure}

\begin{figure}[b!]
	\centering
	\includegraphics[width=8.5cm]{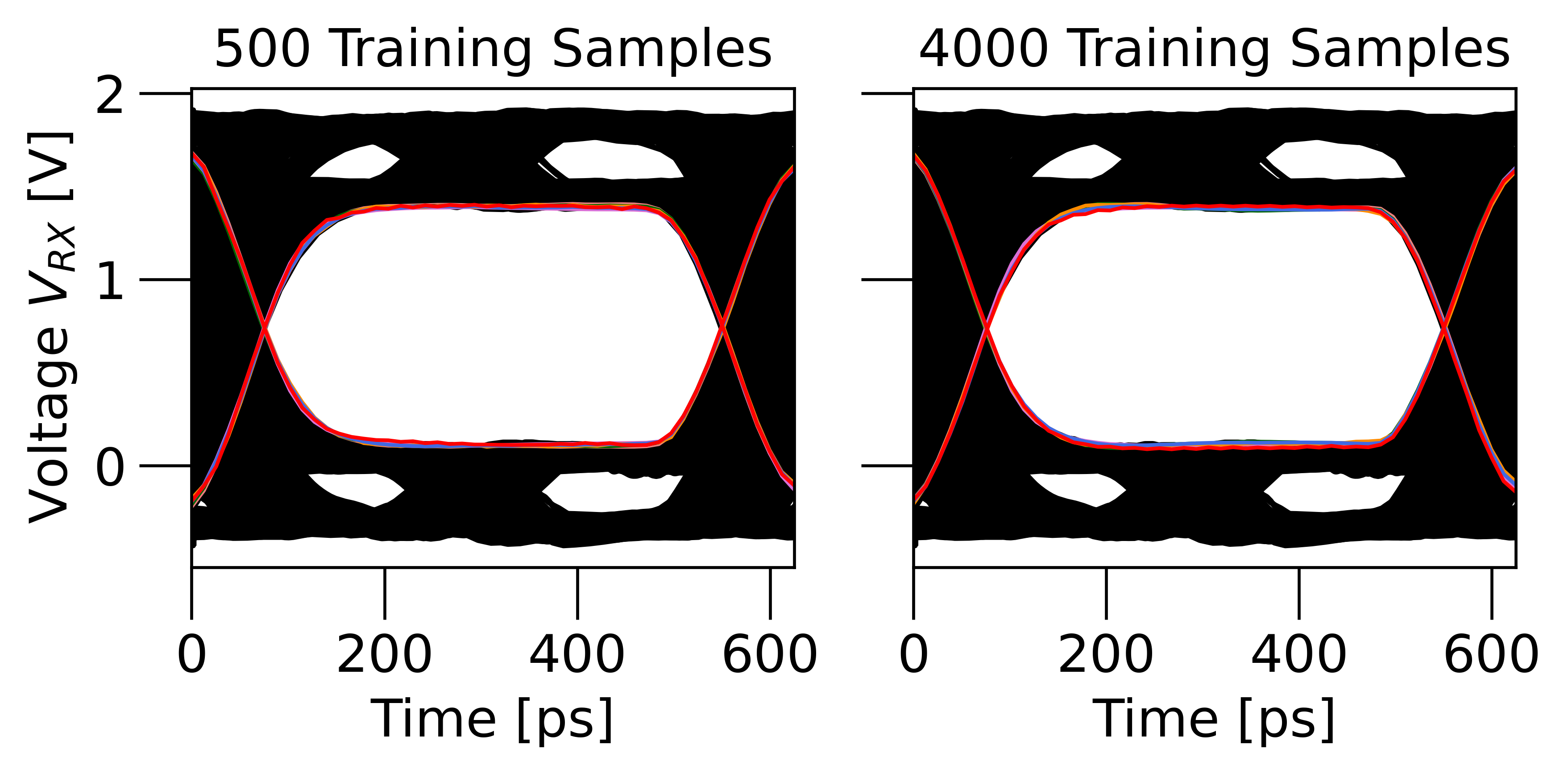}
	\caption{Eye contour predictions of various ML models on a single test sample, with models trained on 500 and 4,000 samples, for the simple fixed-buffer scenario shown in Fig.~\ref{fig_simple_topology}. For color-coding see Fig.~\ref{fig_ml_results_simple_spice}.}
	\label{fig_eye_simple}
\end{figure}

\section{Simple Fixed-Buffer Example}

Initially, the simple fixed-buffer example based on the interconnect shown in Fig.~\ref{fig_simple_topology} was utilized to compare the ML models for prediction of the EH and EW as well as the eye contour. The testing accuracy was evaluated using the root mean squared error (RMSE) and $R^2$-score for a varying number of training sample sizes to analyze the capability of the ML models for different data scarcities. For the eye contour specifically, these metrics were calculated globally across the entire 100-dimensional coordinate vector. The results in Fig.~\ref{fig_ml_results_simple_spice} show that in the low-data regime with 1000 training samples and below, the anisotropic versions of the GPR and KRR are particularly strong, while in the region with more training samples neural networks dominate as expected. In summary, despite differences in performance across data regimes, all ML models demonstrated a strong ability to learn the underlying nonlinear relationship, as their accuracies are closely clustered overall.

This observation can be further supported based on the eye contour prediction example shown in Fig.~\ref{fig_eye_simple}. Predictions from a selection of models for a single test sample are displayed for training sets of 500 and 4000 samples. It is evident that all of the models fit the eye contour well with both training data sizes. This visual evidence is consistent with the quantitative results from Fig.~\ref{fig_ml_results_simple_spice} indicating that the complexity of this task is moderate and therefore easily learnable even with only 500 training samples.

\section{Buffer-Parameterized Approach}

\begin{figure}[b!]
	\centering
	\includegraphics[width=8.5cm]{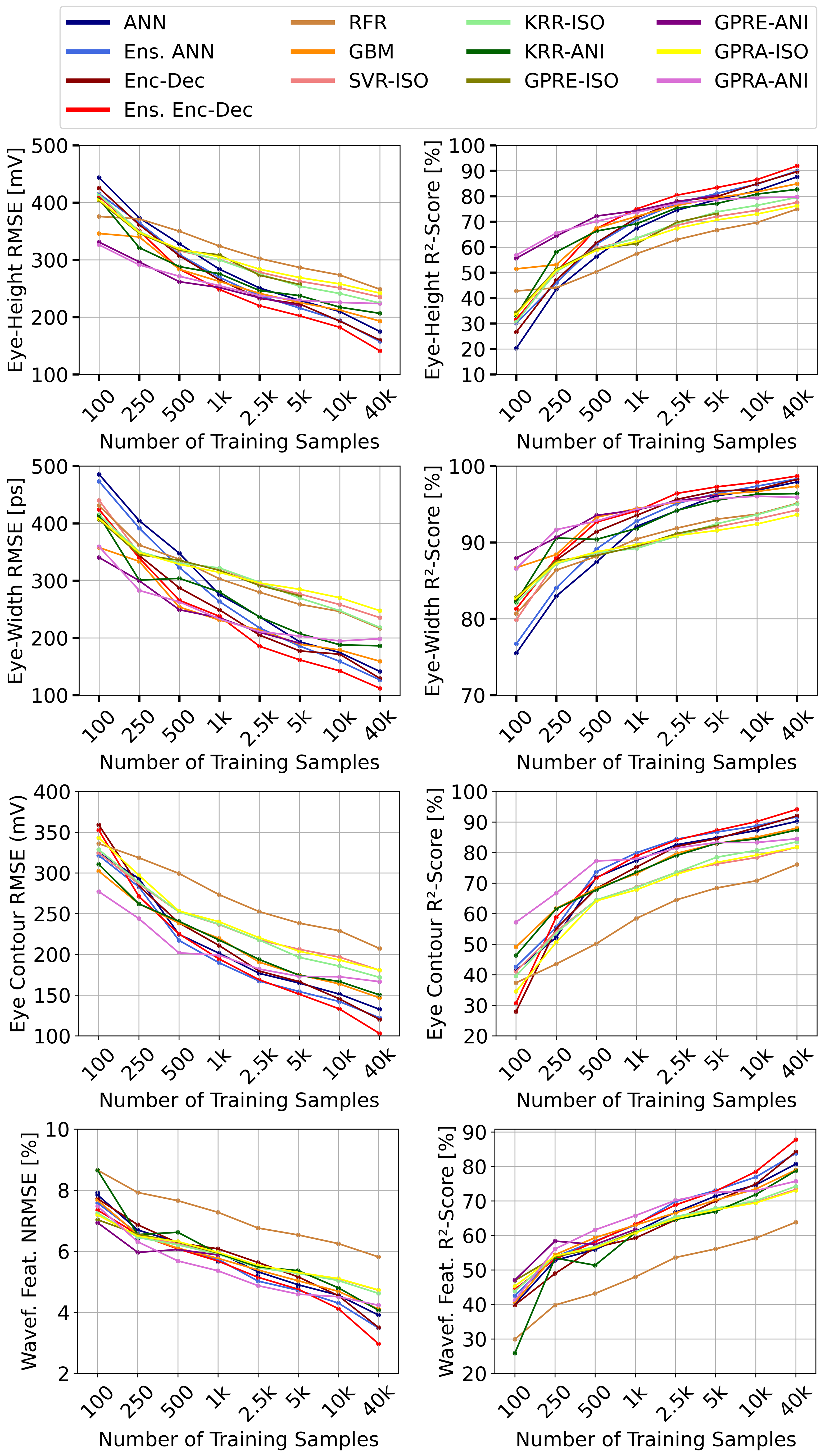}
	\caption{Testing RMSE and $R^2$-score for prediction of EH, EW, eye contour, and transient waveform features for a varying number of training samples based on the buffer-parameterized approach for the point-to-point topology from \cite{Withoeft2025}.}
	\label{fig_ml_results}
\end{figure}

For the buffer-parameterized approach, the point-to-point structure and parameter variation of \cite{Withoeft2025} was utilized. This leads to a higher number of variable PCB parameters with three transmission line lengths and characteristic impedances, the termination resistor, and the dielectric behavior of the PCB material. Additionally, according to the buffer-parameterized approach, the IC buffer parameters for TX and RX were also varied with respect to different clock speeds, voltage levels, rise/fall times, and jitter as well as the internal resistor and capacitor values. This significantly complicates the ML modeling problem as more complex effects and interdependencies have to be taken into account. The testing RMSEs and $R^2$-scores for a varying number of training samples are shown for the prediction of EH, EW, eye contour, and waveform features in Fig.~\ref{fig_ml_results}. The patterns that were only subtly observed in the fixed-buffer modeling scenario are intensified. For larger numbers of training samples the neural networks dominate more significantly, while in low data regimes the anisotropic GPR heavily outperforms the other models. The anisotropic KRR and the GBM offer respectable performance, but resulted in comparatively lower accuracies across all training sample sizes in this specific comparison. For the neural networks, ensembling has a strong positive impact and the encoder-decoder architecture shows its strengths in the region with a maximum number of training samples. Kernel methods and GPR with isotropic kernels as well as RFR performed significantly worse compared to the other models, representing the lower tier of predictive accuracy in this comparison.

\begin{figure}[b!]
	\centering
	\includegraphics[width=8.5cm]{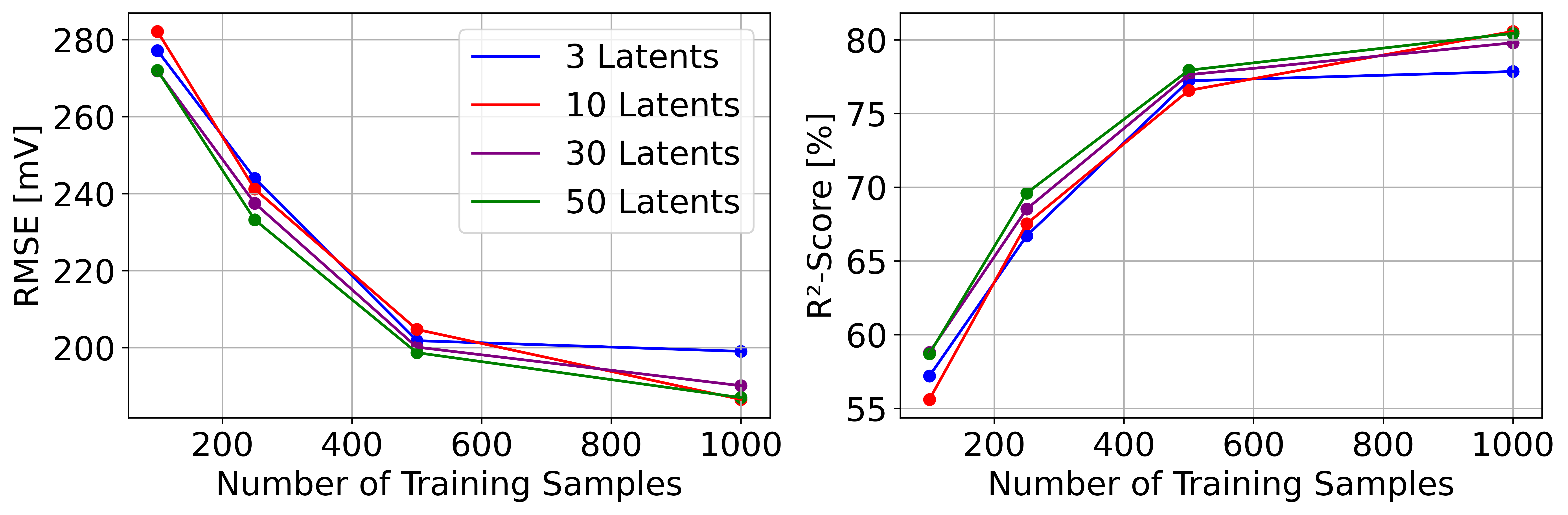}
	\caption{Testing RMSE and $R^2$-score for prediction of the eye contour with an anisotropic approximate GPR model using varying numbers of LCM latents.}
	\label{fig_gpr_contour_detailed}
\end{figure}

Specifically for the prediction of the eye contour using the anisotropic GPR model, different latents of the LCM output kernel were explored in Fig.~\ref{fig_gpr_contour_detailed} to analyze the impact of a higher number of latents compared to the default of 3. While increasing the number of latents can slightly improve the prediction performance, the accuracy with 3 latents is already very close. Moreover, these larger latents also significantly increase computational complexity and make training on larger datasets infeasible even with approximate GP.

\begin{figure}[t!]
	\centering
	\includegraphics[width=8.5cm]{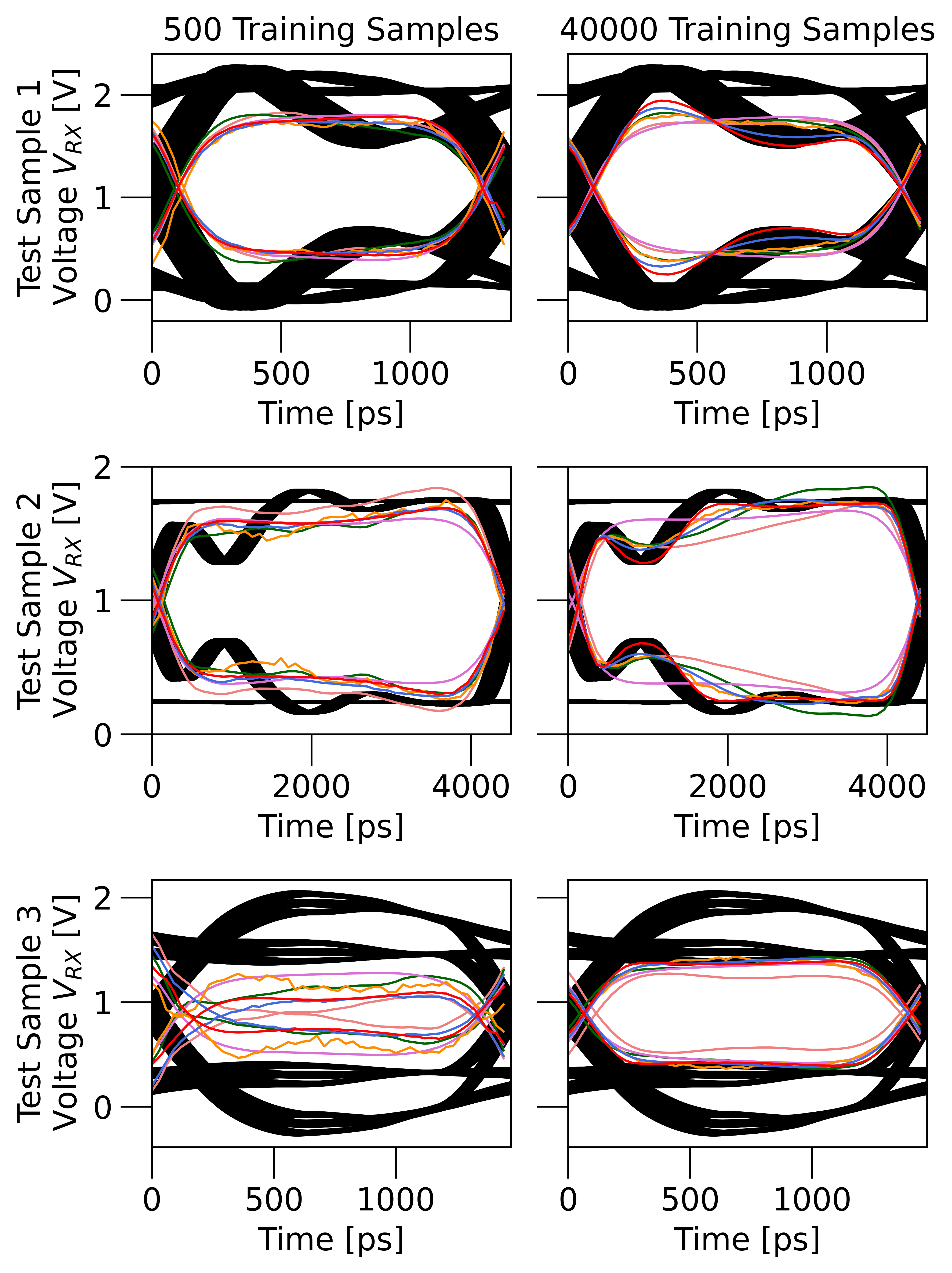}
	\caption{Eye contour predictions of various ML models on three test samples, with models trained on 500 and 40,000 samples, using the buffer-parameterized approach for the point-to-point topology from \cite{Withoeft2025}. For color-coding see Fig.~\ref{fig_ml_results}.}
	\label{fig_eye}
\end{figure}

In Fig.~\ref{fig_eye} some eye contour predictions on the testing dataset are shown based on 500 and 40,000 training samples. The first two test examples demonstrate the superiority of the ensemble encoder-decoder model (red color) trained on 40,000 training samples compared to all of the other models, which learn a representation that is much further from the actual eye contour shape, while with 500 samples all models perform poorly. In the third test example, where the eye contour follows a less complex temporal pattern, it can be observed that except for the SVR all models achieve a good fit if trained with 40,000 training samples. At 500 samples, the anisotropic GPR achieves the best relative accuracy compared to the other models, which is consistent with Fig.~\ref{fig_ml_results}. However, its absolute prediction is still noticeably far from the ground truth. Overall, it is clear that with only 500 training samples unlike the fixed-buffer example the accuracy is insufficient and the increased number of samples is strongly beneficial to achieve good prediction performance, due to the much higher complexity of the modeling task at hand.

\section{Complex Interconnect with added Vias and Package Parasitics}

\begin{figure}[b!]
	\centering
	\includegraphics[width=8.5cm]{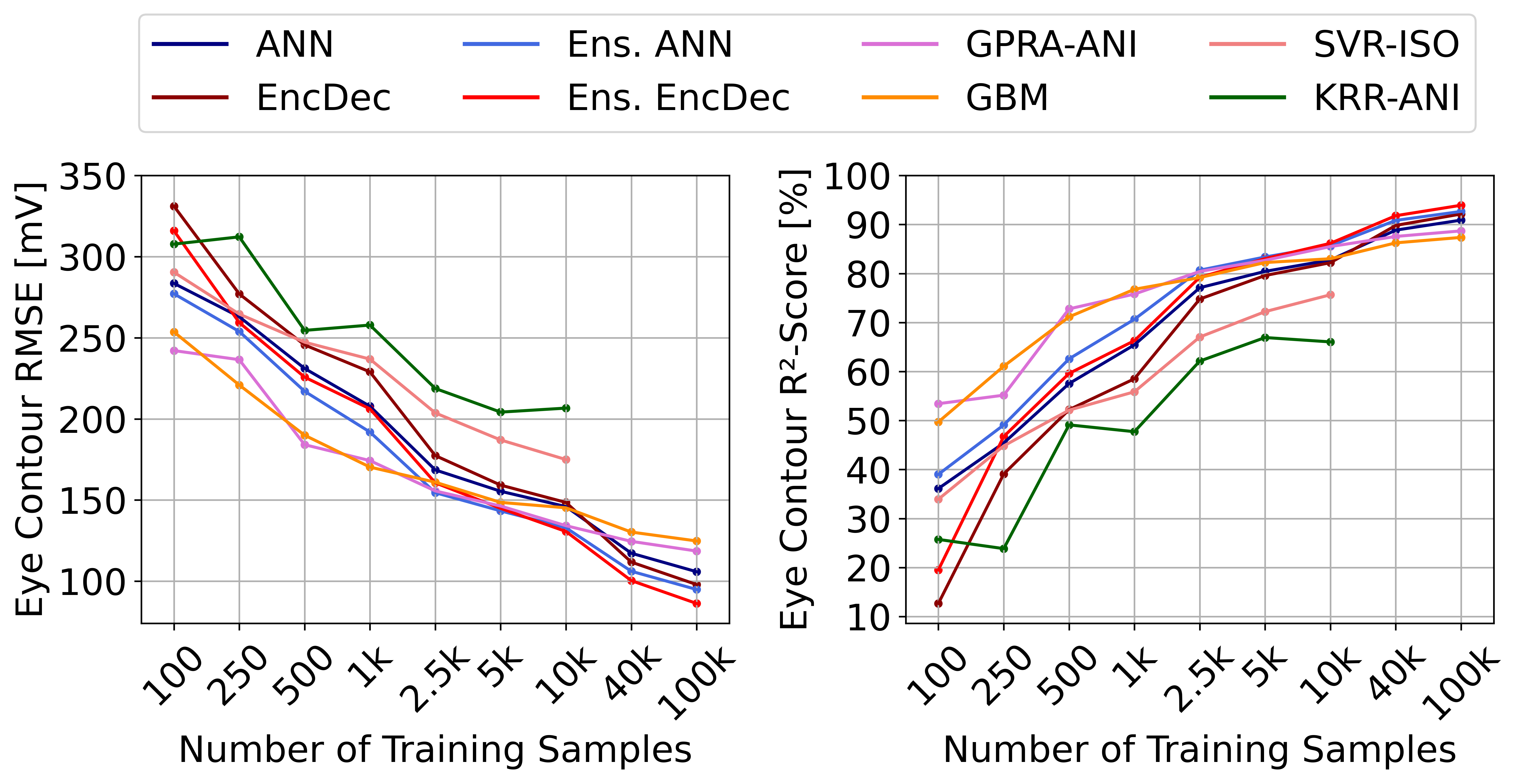}
	\caption{Testing RMSE and $R^2$-score for prediction of the eye contour for a varying number of training samples based on the complex buffer-parameterized scenario with added vias and package parasitics of Fig.~\ref{fig_complex_topology}.}
	\label{fig_eye_complex}
\end{figure}

Finally, for the complex buffer-parameterized scenario shown in Fig.~\ref{fig_complex_topology} with a total of 44 design parameters, the modeling problem becomes even more difficult. Here, the main focus was on eye contour prediction for a selection of models that performed reasonably well in the previous scenario. The testing RMSE and $R^2$-scores for a varying number of training samples are shown in Fig.~\ref{fig_eye_complex}. For this scenario, the anisotropic KRR model degraded significantly due to the high number of design parameters each with a length scale that needs to be tuned through hyperparameter optimization, which becomes infeasible. The SVR model also performed poorly, which is consistent with previous investigations. The GBM and anisotropic GPR models were strong in the low data regime, while the neural networks achieved the best performance only for very large training datasets. This behavior mirrors the observations from the buffer-parameterized scenario, with GPR and GBM outperforming neural networks at small to moderate data sizes, but a clear shift toward a more data-intensive regime is evident. This can be attributed to the larger number of design parameters.

\begin{table}[t!]
	\caption{Model performance comparison based on the $R^2$-score and the sample-wise NRMSE normalized by the maximum vertical swing of the inner eye contour}
	\centering
	\label{tab_model_errors_norm}
	\begin{tabular}{lcccccc}
		\toprule
		\textbf{Model} & \multicolumn{3}{c}{\textbf{NRMSE [\%]}} & \multicolumn{3}{c}{\textbf{$\mathbf{R^2}$-Score [\%]}} \\
		\cmidrule(lr){2-4} \cmidrule(lr){5-7}
		\textbf{Fig.~\ref{fig_ml_results}} & \textbf{10k} & \textbf{40k} & \textbf{100k} & \textbf{10k} & \textbf{40k} & \textbf{100k} \\
		\midrule
		Enc-Dec               & 8.7  & 6.6  & -- & 88.2 & 91.9 & -- \\
		Ensemble Enc-Dec      & 7.6  & 5.6  & -- & 90.1 & 94.1 & -- \\
		ANN                   & 9.5  & 8.2  & -- & 87.3 & 90.2 & -- \\
		Ensemble ANN          & 8.5  & 7.2  & -- & 88.7 & 91.7 & -- \\
		GPR Anisotropic       & 11.4 & 10.8 & -- & 83.3 & 84.5 & -- \\
		GBM                   & 10.0 & 8.9  & -- & 85.0 & 88.0 & -- \\
		\midrule
		\textbf{Fig.~\ref{fig_eye_complex}} & \textbf{10k} & \textbf{40k} & \textbf{100k} & \textbf{10k} & \textbf{40k} & \textbf{100k} \\
		\midrule
		Enc-Dec               & 12.2 & 8.8  & 7.7 & 82.2 & 89.8 & 92.1 \\
		Ensemble Enc-Dec      & 10.2 & 7.7  & 6.6 & 86.2 & 91.8 & 93.9 \\
		ANN                   & 11.6 & 9.4  & 8.5 & 82.8 & 88.9 & 90.9 \\
		Ensemble ANN          & 10.6 & 8.3  & 7.5 & 85.7 & 90.8 & 92.7 \\
		GPR Anisotropic       & 11.3 & 10.3 & 9.9 & 85.5 & 87.6 & 88.7 \\
		GBM                   & 11.8 & 10.4 & 9.8 & 83.0 & 86.3 & 87.4 \\
		\bottomrule
	\end{tabular}
\end{table}

\begin{figure}[b!]
	\centering
	\includegraphics[width=8.5cm]{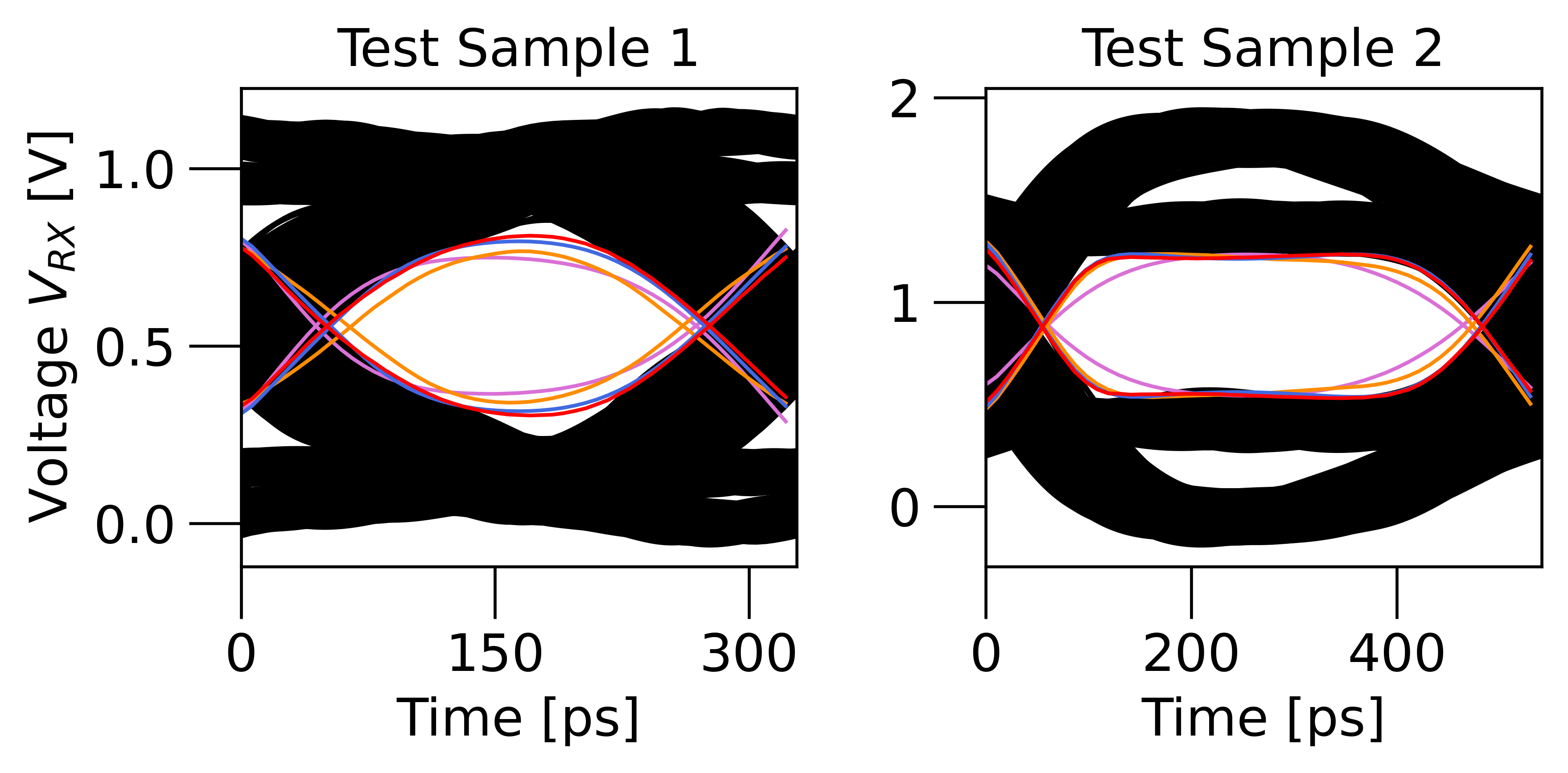}
	\caption{Eye contour predictions of various ML models on two test samples, with models trained on 100,000 samples, for the complex buffer-parameterized scenario including vias and package parasitics shown in Fig.~\ref{fig_complex_topology}. For color-coding see Fig.~\ref{fig_eye_complex}.}
	\label{fig_complex_contour}
\end{figure}

In general, increasing the number of training samples is necessary to achieve comparable performance, due to the higher modeling complexity introduced by the significantly larger design parameter dimensionality. This is also backed by the data shown in Table \ref{tab_model_errors_norm}, which compares the sample-wise NRMSE, normalized by the maximum vertical swing of the inner eye contour, and the $R^2$-scores of the best-performing models for the eye contour prediction task of Fig.~\ref{fig_ml_results} and Fig.~\ref{fig_eye_complex}. The results indicate that for roughly the same prediction accuracy, 2.5 times the training data was needed for the complex case. This approximately corresponds to the threefold increase in design parameter dimensionality, which indicates that within the investigated range, the required amount of training data increases nearly proportionally with the number of design parameters.

Fig.~\ref{fig_complex_contour} shows the eye contour predictions on the test dataset using the ensemble ANN, ensemble encoder-decoder, GBM, and anisotropic GPR models trained with 100,000 samples. It is visible that all models achieved high prediction accuracy for both test samples, with the neural networks showing a slight advantage in both cases over GPR and GBM, which aligns with the observations in Fig.~\ref{fig_eye_complex}.

\section{Computational Efficiency and Acceleration}

Before the buffer-parameterized framework can be used for design space exploration and optimization, its computational efficiency has to be evaluated to justify the initial data generation phase. Table \ref{tab_runtime_detailed} summarizes the computational cost required for the complex 44 design parameter topology. To create the surrogate model; a significant one-time investment is required. This includes the generation of 120,000 total simulation samples for the training, validation, and testing datasets, which takes 53 hours on a CPU. The data generation process is followed by 19.94 minutes of GPU-accelerated training for the most accurate encoder-decoder model.

\begin{table}[b!]
	\caption{Detailed computational cost summary for data generation, training, and inference for the complex 44 design parameter topology}
	\label{tab_runtime_detailed}
	\centering
	\setlength{\tabcolsep}{4pt} 
	\begin{tabular}{@{}llccc@{}}
		\toprule
		\textbf{Phase} & \textbf{Method} & \textbf{Samples} & \textbf{Runtime} & \textbf{Speedup} \\
		\midrule
		\multicolumn{5}{c}{\textit{Initial Model Development (One-Time Investment):}} \\
		Train Data Gen. & Simulation [CPU] & 100,000 & 44.17 h & -- \\
		Test Data Gen. & Simulation [CPU] & 10,000 & 4.47 h & -- \\
		Val. Data Gen. & Simulation [CPU] & 10,000 & 4.36 h & -- \\
		\midrule
		Total Data Gen. & Simulation [CPU] & 120,000 & 53.00 h & -- \\
		\midrule
		Model Training & Enc-Dec [GPU] & 100,000 & 19.94 m & -- \\
		\midrule
		\multicolumn{5}{c}{\textit{Design Space Exploration \& Optimization:}} \\
		Inference & Simulation [CPU] & 500,000 & 220.83 h & 1$\times$ \\
		Inference & Enc-Dec [CPU] & 500,000 & 135.43 s & $\approx5,871\times$ \\
		Inference & Enc-Dec [GPU] & 500,000 & 4.98 s & $\approx159,600\times$ \\
		\bottomrule
	\end{tabular}
	
	\vspace{2pt}
	\parbox{\linewidth}{\centering\scriptsize\textit{Hardware:} AMD Ryzen 9 7900, NVIDIA RTX 4080 Super, 64 GB DDR5 RAM}
\end{table}

However, the advantage of the buffer-parameterized approach lies in the amortization of this initial cost through model reusability. The trained surrogate framework can be applied to any arbitrary IC buffer setting or corner case within the defined parameter bounds without ever requiring a new simulation or retraining cycle. To demonstrate the value of this reusability, the framework was used for four design scenarios including varying DDR generations and impedance tolerances and a large scale of 500,000 parameter samples. Utilization of traditional CPU-based simulation to execute these 500,000 runs would require over 220 hours of continuous computation.

The surrogate model runs inference on the entire 500,000 parameter set in 135.43 seconds on a CPU resulting in an approximate 5,871$\times$ speedup over simulation. When hardware acceleration is utilized, GPU-based inference completes the task in 4.98 seconds, resulting in an approximate 159,600$\times$ speedup over simulation. While a single, narrow optimization run might be faster using direct simulation, cross-technology workflows for practical SI design space exploration and optimization require a large number of iterative evaluations. Because the inference using the surrogate model is orders of magnitude faster than simulation, the proposed methodology transforms large-scale design space exploration into a very efficient verification process.

\section{Design Space Exploration using the Buffer-Parameterized Machine Learning Models}

To further demonstrate the applicability and scalability of the buffer-parameterized ML framework, design optimization scenarios targeting the fulfillment of specific eye mask requirements were investigated. These masks and the corresponding IC buffer settings, extracted from IBIS models and datasheets, are detailed in Tables \ref{tab_ddr_eye_masks} and \ref{tab_ddr_buffer_settings}, respectively. The eye contour ML models can be utilized for compliance evaluation based on these targets for DDR3, DDR4, and DDR5 technologies. However, these investigated settings are only representative demonstration cases. The framework can be applied to any arbitrary buffer characteristics or eye mask constraints without changing the underlying ML architecture.

\begin{table}[b!]
	\caption{Utilized eye mask settings}
	\centering
	\setlength{\tabcolsep}{4pt}
	\begin{tabular}{lcccc}
		\toprule
		Speedgrade & UI [ps] & $V_{\text{ref}}$ [V] & $t_0$/$t_1$ [ps] & $V_0$/$V_1$ [V] \\
		\midrule
		DDR3-1600 & 625 & 0.75 & 174 / 469 & 0.55 / 0.95 \\
		DDR4-1600 & 625 & 0.60 & 174 / 469 & 0.50 / 0.70 \\
		DDR5-4800 & 208 & 0.55 & 58 / 156 & 0.50 / 0.60 \\
		\bottomrule
	\end{tabular}
	\label{tab_ddr_eye_masks}
\end{table}

\begin{table}[b!]
	\caption{Utilized IC buffer settings}
	\centering
	\setlength{\tabcolsep}{4pt}
	\begin{tabular}{lccccc}
		\toprule
		Speedgrade & $V_{\text{dd}}$ [V] & $t_{\text{r}} / t_{\text{f}}$ [ps] & $R_{\text{tx}}$ [$\Omega$] & $C_{\text{tx}}$ [pF] & $C_{\text{rx}}$ [pF] \\
		\midrule
		DDR3-1600 & 1.5 & 320 & 32 & 4.3 & 0.7 \\
		DDR4-1600 & 1.2 & 115 & 32 & 0.78 & 0.43 \\
		DDR5-4800 & 1.1 & 50 & 30 & 1.0 & 1.0 \\
		\bottomrule
	\end{tabular}
	\label{tab_ddr_buffer_settings}
\end{table}

\subsection{DDR3 Scenario}

\begin{figure}[t!]
	\centering
	\includegraphics[width=8.5cm]{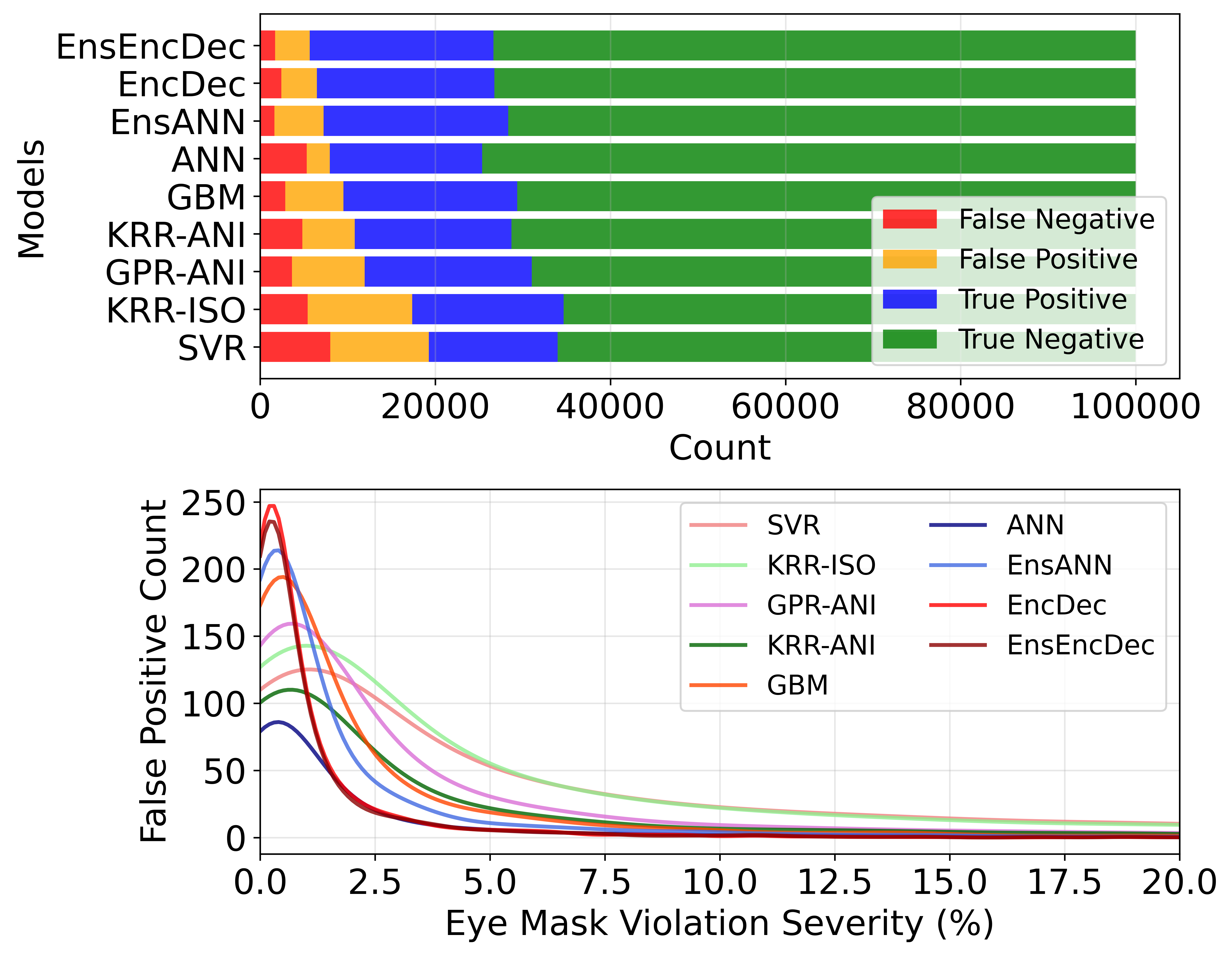}
	\caption{Statistical evaluation of the DDR3 design optimization showing eye mask classification and true positive violation severity per ML model.}
	\label{fig_ddr3_statistics}
\end{figure}

Firstly, for the DDR3 scenario, 100,000 samples were generated using LHS within the relevant parameter space, while the remaining parameters were fixed based on the buffer and impedance settings (initially 50~$\Omega$ matched). The ML model predictions regarding eye mask compliance (0 = not met, 1 = met) were then statistically compared to the ground-truth simulation results in Fig.~\ref{fig_ddr3_statistics}. As expected, the previously best-performing models on the testing dataset achieved the lowest number of classification errors. Additionally, in the case of false positives, the severity of the eye mask violations, which is defined as the percentage of the total mask area in $\text{mV} \times \text{ps}$ encroached by the simulated inner eye contour, showed a significantly tighter statistical distribution for the more accurate ML models. As detailed by the density curves in Fig.~\ref{fig_ddr3_statistics}, the optimal neural network architecture has most of its false positive violations near 0\% and the error distributions tail off almost completely before the 5\% mark.  In contrast, baseline models such as the isotropic KRR and SVR exhibited heavy-tailed distributions, with violation severities spreading well beyond 15\%. While it was possible to reduce the number of false positives by extending the mask voltages beyond the required eye mask values to essentially shift the decision threshold, this inevitably led to a heavy increase in false negatives.

\begin{figure}[b!]
	\centering
	\includegraphics[width=8.5cm]{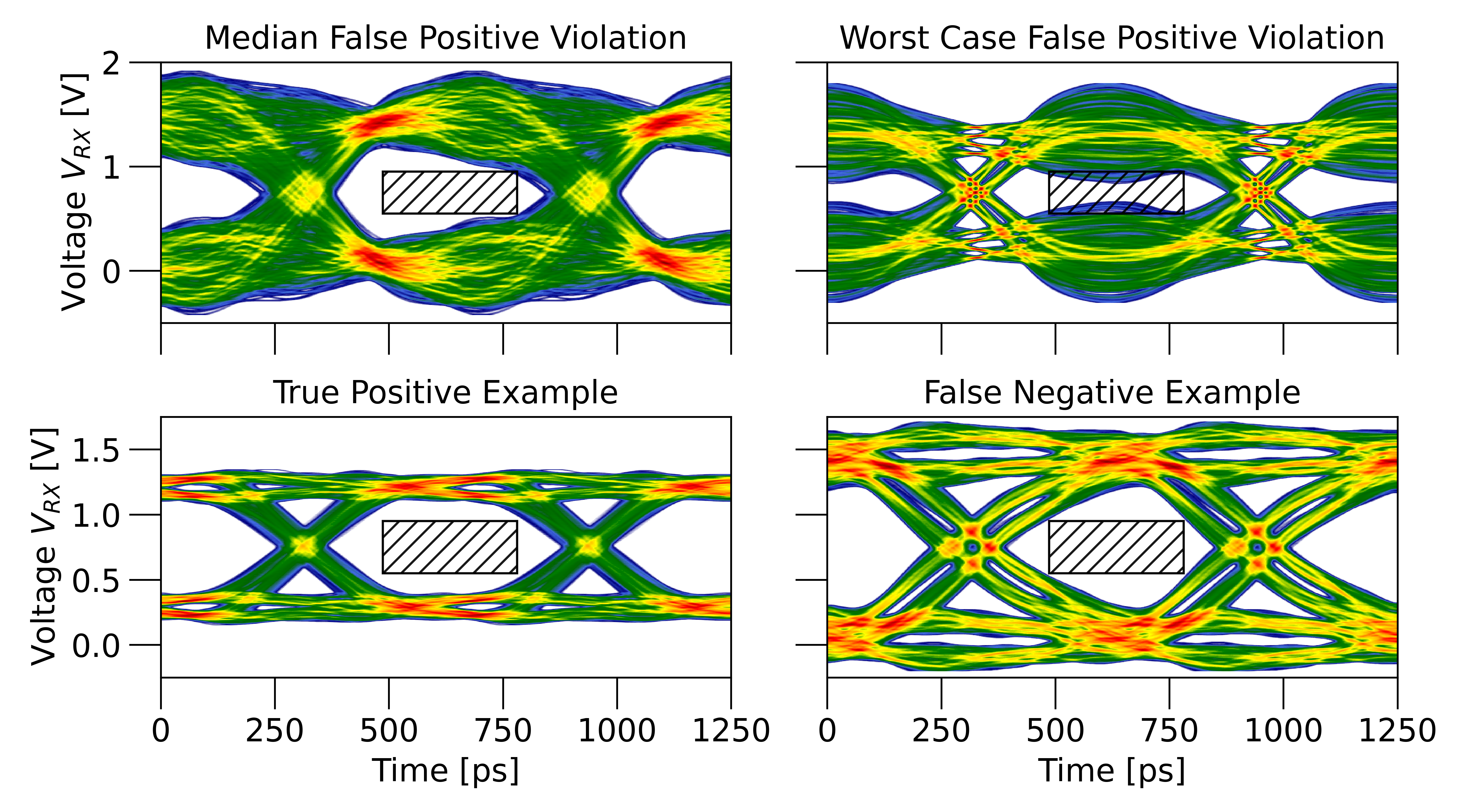}
	\caption{Eye diagram examples for the DDR3 design scenario optimization with a matched characteristic impedance of 50~$\Omega$.}
	\label{fig_eye_50}
\end{figure}

Fig.~\ref{fig_eye_50} shows representative eye diagrams based on predictions of the best-performing encoder-decoder model. The median false positive violation example exhibited only a slight infringement of the eye mask at the right edge, indicating that most violations were minor and occurred close to the decision boundaries rather than representing gross errors. The worst-case false positive violation was more pronounced but remained within a reasonable range, as it corresponded to an edge case in which the eye opening was still largely maintained. These representative cases demonstrate that model errors are not arbitrary but are concentrated in explainable edge scenarios relative to the mask. The true positive example represented the most frequent case, with a well-defined eye opening that fit the mask with adequate margins, reflecting the model's ability to correctly identify compliant designs. The false negative example corresponded to an eye diagram that was close to the mask limits, making the misclassification plausible. Overall, these visualizations confirm that violations and misclassifications predominantly occurred under marginal eye mask conditions, rather than indicating significant deficiencies of the model.

\begin{figure}[t!]
	\centering
	\includegraphics[width=8.5cm]{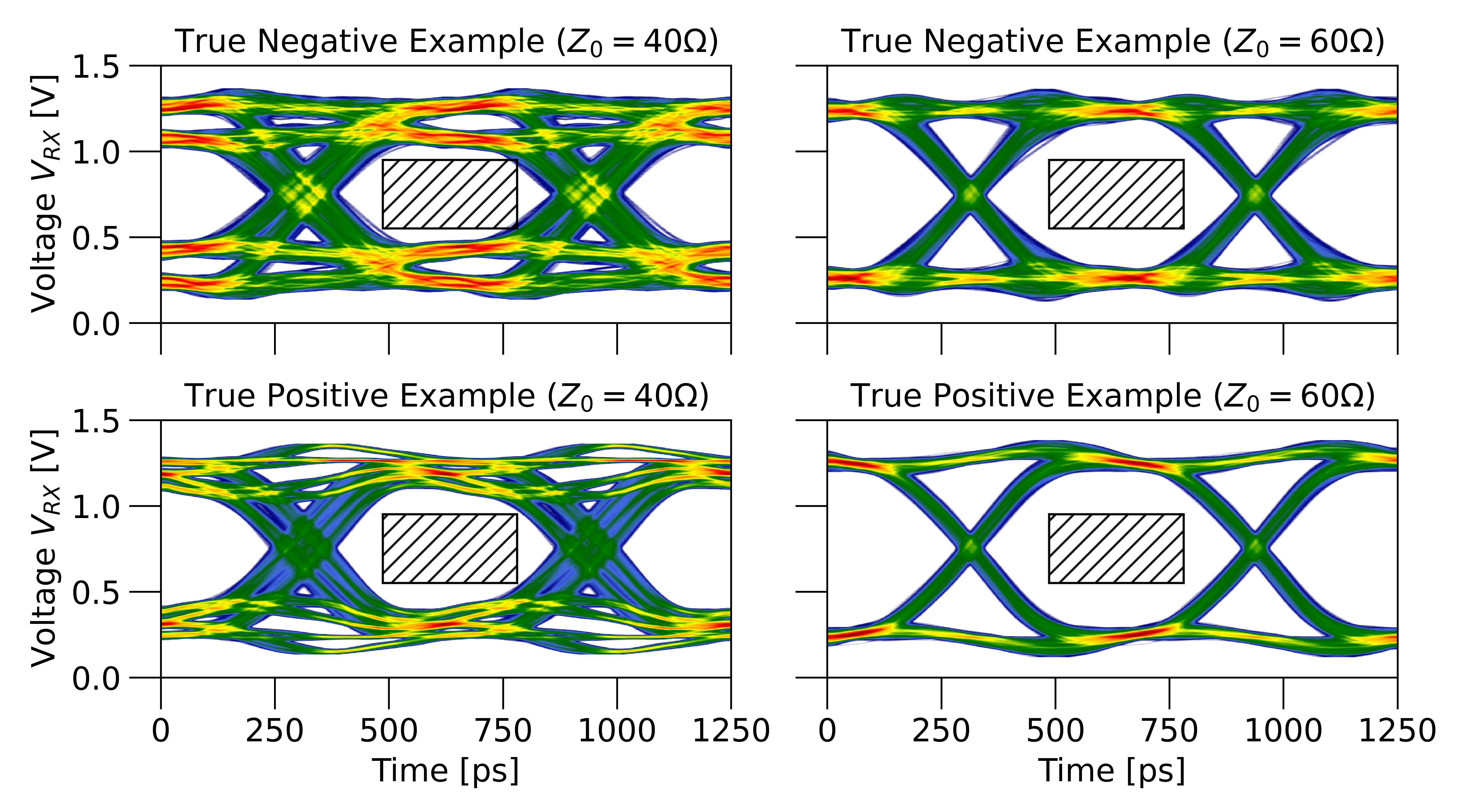}
	\caption{Eye diagram examples for the DDR3 design scenario optimization based on a nominal characteristic impedance of 50~$\Omega$ and manufacturing tolerances of $\pm 20\%$ leading to 40~$\Omega$ (left) and 60~$\Omega$ (right).}
	\label{fig_eye_40_60}
\end{figure}

\subsection{Validation Under Manufacturing Tolerances}

\begin{figure}[b!]
	\centering
	\includegraphics[width=8.5cm]{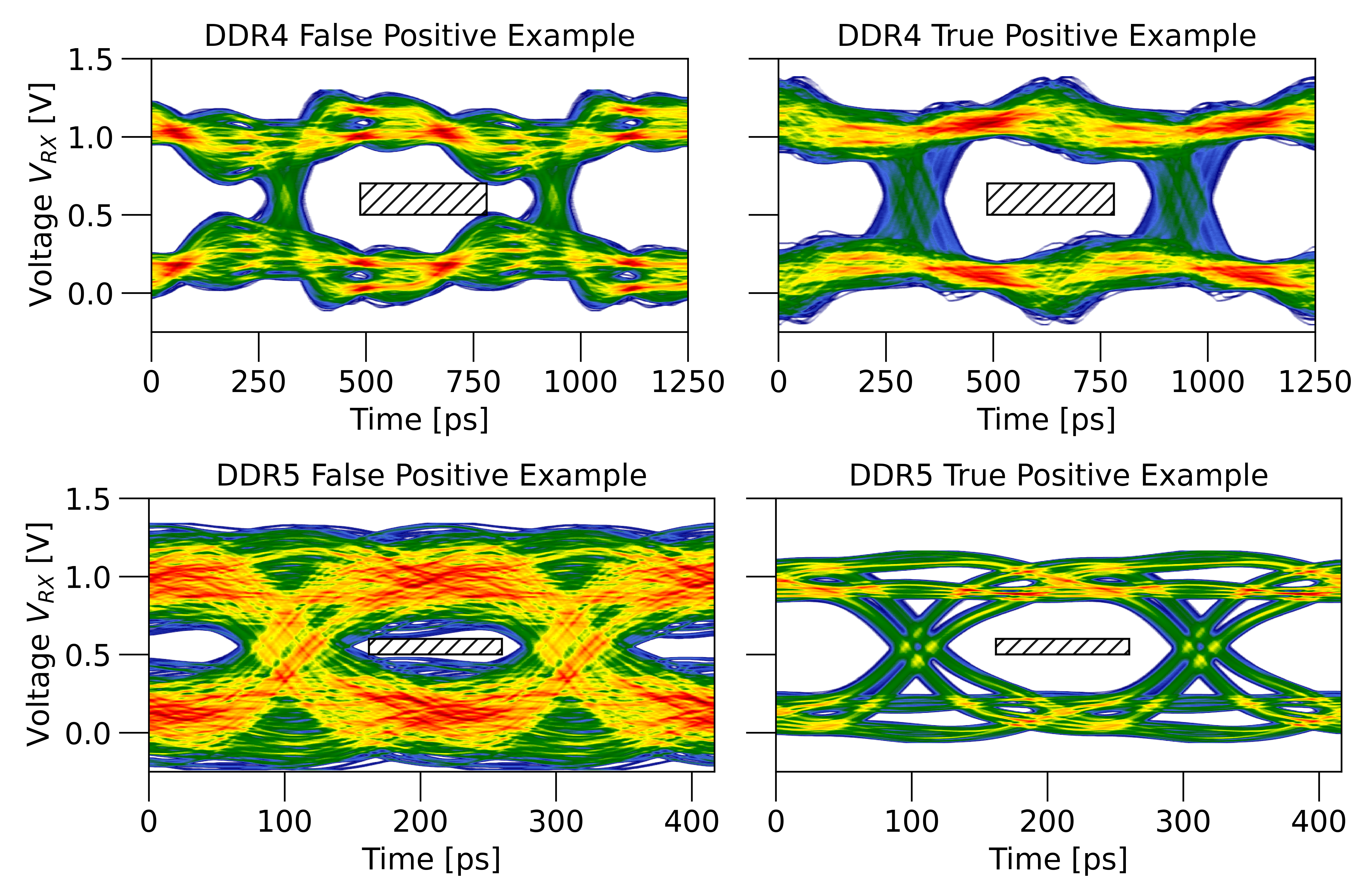}
	\caption{Eye diagram examples for the DDR4 (top) and DDR5 (bottom) design scenario optimization with a matched characteristic impedance of 50~$\Omega$.}
	\label{fig_eye_ddr45}
\end{figure}

Secondly, tolerances with respect to the characteristic impedance were introduced to create a more realistic setting. For a nominal characteristic impedance of 50~$\Omega$ and a tolerance of $\pm 20\%$, the corresponding corner cases are 40~$\Omega$ and 60~$\Omega$. The same LHS setup as described previously was therefore generated twice, once with a characteristic impedance of 40~$\Omega$ and once with 60~$\Omega$. Based on these datasets, ML model predictions were obtained, while the ground truth eye was obtained via simulation. A design was considered valid only if the eye mask was met for both 40~$\Omega$ and 60~$\Omega$. As expected, the number of valid designs was drastically reduced from previously 22,702 to 4,883, which was also reflected in the ML model predictions. Among all tested models, the encoder-decoder model achieved the highest accuracy again and was thus utilized for further analysis of the eye diagrams. This analysis revealed that the design that previously passed at 50~$\Omega$ as the true positive example was now correctly declared by the model as invalid, since at 40~$\Omega$ a small but noticeable eye mask violation occurred as shown in Fig.~\ref{fig_eye_40_60}. Another design that had similar characteristics but did not violate the mask at 40~$\Omega$ was also correctly classified as valid. These observations indicate that even small differences in the eye contour are well captured by the buffer-parameterized ML model. Overall, the results demonstrate the applicability of the proposed approach under varying conditions, such as deviations from the nominal 50~$\Omega$ characteristic impedance due to manufacturing tolerances.

\subsection{Cross-Technology Extension to DDR4/5}

Finally, the DDR4 and DDR5 eye mask and buffer characteristics were analyzed. Fig.~\ref{fig_eye_ddr45} shows representative false and true positive eye predictions from the encoder-decoder model. For the DDR5 case the complex buffer-parameterized ML model of Fig.~\ref{fig_eye_complex} was used. The results show that for both DDR4 and DDR5, the model correctly classifies passed and failed eye masks. This analysis confirms that the model is accurate even when different buffer settings are considered. Errors are confined to designs near the mask boundaries, which demonstrates the adaptability of the approach under different technological settings.

\section{Conclusion and Future Work}

This paper demonstrates the effectiveness of buffer-parameterized ML approaches for SI analysis and optimization that include buffer characteristics as trainable input parameters. This approach offers significant advantages over fixed-buffer methods, as it can adapt to varying buffer configurations without retraining. It was also shown that the complexity of the interconnect and the number of design parameters influence the amount of training data that is required to achieve accurate predictions. Neural networks outperform other ML models when trained on large datasets, whereas anisotropic GPR achieves superior performance in low-data scenarios. These findings highlight the potential of buffer-parameterized ML-based solutions for SI analysis, design space exploration, and optimization, which was demonstrated using scenarios with varying eye mask requirements and different DDR generations to identify valid and non-compliant designs.

Future research will extend the buffer-parameterized approach to model more complex SI effects such as equalization, highly non-linear buffer behavior, as well as coupled and differential transmission lines. Discrete equalization settings, such as FFE or DFE tap coefficients, can be integrated as additional input parameters. The characteristics of non-linear IBIS profiles can be compressed into learnable embeddings using dimensionality reduction to circumvent the curse of dimensionality. Furthermore, the framework can be extended to differential signaling and coupled transmission lines through the incorporation of trace spacings as trainable input parameters to include crosstalk and mode-conversion effects.
	
To overcome the curse of dimensionality and scale to these larger parameter spaces, the data generation should be focused on active learning methodologies. From a modeling perspective, graph neural networks (GNNs) could encode the connectivity patterns of interconnect structures to improve generalization with respect to different topologies such as fly-by, star, and point-to-point. The combination of these structural representations with the buffer-parameterized approach within an encoder-decoder architecture would result in a framework for multi-task and meta-learning. This strategy would enable encoders to adapt to different topologies and design conditions, and decoders to specialize in the prediction of various SI metrics to support application for different design scenarios.

\bibliography{paper_bib}

\bibliographystyle{IEEEtran}

\end{document}